# Strong Spin Hall Effect in the Antiferromagnet PtMn

Yongxi Ou[1], Shengjie Shi[1], D. C. Ralph[1,2], and R. A. Buhrman[1,*]

[1]Cornell University, Ithaca, New York 14853, USA

[2]Kavli Institute at Cornell, Ithaca, New York 14853, USA

Abstract

Effectively manipulating magnetism in ferromagnet (FM) thin film nanostructures with an in-plane current has become feasible since the determination of a "giant" spin Hall effect (SHE) in certain heavy metal (HM)/FM system. Recently, both theoretical and experimental reports indicate that the non-collinear and collinear metallic antiferromagnet (AF) materials can have both a large anomalous Hall effect (AHE) and a strong SHE. Here we report a systematic study of the SHE in PtMn with several PtMn/FM systems. By using interface engineering to reduce the "spin memory loss" we obtain, in the best instance, a spin torque efficiency

$$\xi_{DL}^{PtMn} \equiv T_{int}\theta_{SH}^{PtMn} \simeq 0.24,$$ where $T_{int}$ is the effective interface spin transparency. This is more than twice the previously reported spin torque efficiency for PtMn. We also find that the apparent spin diffusion length in PtMn is surprisingly long, $\lambda_s^{PtMn} \approx 2.3\mathrm{nm}$.

SHE in different heavy metal (HM)/ferromagnet (FM) systems[1–4] can be characterized by the spin Hall ratio (angle) $\theta_{SH} \equiv (2e/\hbar)J_s/J_e$ where $J_s$ is the transverse spin current density generated in the HM and $J_e$ is the applied longitudinal electrical current density. Recently a new class of heavy metal (HM) alloys, the non-collinear antiferromagnet (AF), $Mn_3Ir$[5–7] and Cu-Au-I type AF, $X_{50}Mn_{50}$ (X=Fe, Pd, Ir, and Pt)[8–11] have been reported to exhibit SHE as spin current sources, with an internal $\theta_{SH}^{PtMn} \approx 0.125$ for PtMn[10], opening up a new area in the rapidly advancing field of "antiferromagnet spintronics"[12–17]. To date research on the SHE from AFs has utilized the implicit assumption that there is no interfacial spin flip scattering or "spin memory loss" (SML)[18] when the spin current traverses the interface to apply a torque to the FM. However the existence of a large SML at some Pt/FM interfaces, together with the negative enthalpy of formation of Mn with both Fe and Ni[19] that can promote interface intermixing, raises the question whether there may also be a significant SML at PtMn/FM interfaces, which would mean that the internal $q_{SH}^{PtMn}$ within PtMn could actually be much higher than previously reported.

We performed a systematic study of the SHE in several PtMn/FM systems employing spin-torque ferromagnetic resonance (ST-FMR)[20] on in-plane magnetized (IPM) FM layers and the harmonic response technique (HR)[21,22] on FM layers with perpendicular magnetic anisotropy (PMA). We also studied samples where a thin $(0.25\,nm - 0.8\,nm)$ Hf layer is inserted between the PtMn and the FM to suppress strong SML at the interface[23]. We find $\xi_{DL}$ to vary significantly with both the deposition order for a given PtMn/FM system and between the different FM systems, but to be relatively consistent between IPM and PMA samples with the same constituents. We also obtained robust current-induced switching in these PMA samples demonstrating the potential for utilizing PtMn in perpendicular magnetic tunneling junction (p-MTJ) and three-terminal device applications.

We first fabricated a series of IPM PtMn/Co bilayer samples by sputter deposition (see Methods) for ST-FMR measurement of the *anti-damping* and *field-like* spin torque efficiencies, $\xi_{DL}$ and $\xi_{FL}$. The magnetic properties of the samples were also characterized by vibrating sample magnetometry (VSM). Because the order in which the HM and FM layers are deposited affected $\xi_{DL}$ in a previous Pt/Co study[24], we grew the PtMn/Co multilayers in both the "standard order" (SO) Ta(1.5)/PtMn(8)/Co($t_{Co}$)/MgO(1.6)/Ta(1.5) (series A) and in the "reversed order" (RO) MgO(1.6)/Co($t_{Co}$)/PtMn(8)/MgO(1.6)/Ta(1.5) (series B) (number in parenthesis is thickness in nm). The ST-FMR measurement schematic is illustrated in Fig. 1a. In this technique we obtain the FMR spin torque efficiency $\xi_{FMR}$ that is obtained from the ratio of the symmetric and antisymmetric components of anisotropic magnetoresistance response at the ferromagnetic resonance. The symmetric part is proportional to the *anti-damping* torque and the antisymmetric part is due to the sum of the Oersted field torque and the *field-like* torque (Ref.[24] and Supplementary Information). Fig. 1b shows the results $\xi_{FMR}$ as a function of Co thickness $t_{Co}$ for both the standard (main) and reversed (inset) order samples. For the SO PtMn/Co samples, the spin current in the PtMn layer generates a significant *field-like* torque in addition to the *anti-damping* torque and consequently $\xi_{FMR}$ varies significantly with thickness. By plotting $1/\xi_{FMR}$ *vs.* $1/t_{Co}$, $\xi_{DL}$ can be determined from the $1/t_{Co}=0$ intercept and the *field-like* spin torque efficiency $\xi_{FL}$ can be determined from the slope of the plot, provided $\xi_{FL}$ is effectively independent of $t_{Co}$ (Supplementary Information). For the reversed order Co/PtMn samples $\xi_{FMR}(\approx \xi_{DL})$ is essentially constant *vs.* $t_{Co}$, indicating $\xi_{FL}$ is negligible (Fig. 1a inset). From this we obtain $\xi_{DL}=0.16\pm0.01$ and $\xi_{FL}=-0.040\pm0.008$ for the SO samples and $\xi_{DL}$ (average) $=0.19\pm0.02$ and $\xi_{FL} \sim 0$ for the RO samples (The minus sign for $\xi_{FL}$ indicates that the *field-like* effective field is *opposite* to the Oersted field).

To further confirm this result with another FM material and to examine the PtMn SHE in structures with PMA, which we were not able to obtain with PtMn/Co bilayers, we replaced Co with $Fe_{60}Co_{20}B_{20}$ (FeCoB) for the FM layer. First we fabricated two IPM series of PtMn/FeCoB bilayers samples, a SO series (C): Ta(1.5)/PtMn(8)/FeCoB($t_{FeCoB}$)/MgO(1.6)/Ta(1.5) and a RO series (D): MgO/FeCoB($t_{FeCoB}$)/PtMn(8)/MgO(1.6)/Ta(1.5). In Fig. 1c we show $1/\xi_{FMR}$ vs. $1/t_{FeCoB}^{eff}$ as obtained for these two sets of samples. From the linear fits to the plots we obtained $\xi_{DL} = 0.096 \pm 0.003$, $\xi_{FL} = -0.043 \pm 0.003$ for the SO series (C) samples and $\xi_{DL} = 0.174 \pm 0.004$, $\xi_{FL} = -0.036 \pm 0.002$ for the RO series (D) samples.

We also fabricated a SO series (E) of Ta(1.5)/PtMn(8)/FeCoB($t_{FeCoB}$)/MgO(1.6)/Ta(1.5) with a thinner FM range of $0.4 \text{nm} < t_{FeCoB} < 1.5 \text{nm}$, the mid-range of which exhibited PMA without any high temperature annealing (Supplementary Information). The highest out-of-plane anisotropy field $H_{an} \approx 1.8 \text{kOe}$ was achieved with $t_{FeCoB} \approx 0.8 \text{nm}$, which allows us to perform HR measurement of the efficiency of the spin torques exerted on the perpendicularly magnetized FM. The results were $\xi_{DL} = 0.11 \pm 0.02$ and $\xi_{FL} = -0.04 \pm 0.02$, in accord with the ST-FMR values obtained via ST-FMR from the IPM series (C) samples with the same layer structure but thicker

Recent work[23,25] has shown that the insertion of a thin layer of Hf between FeCoB and the HM in a spin Hall device structure can substantially enhance the PMA, while the thin Hf ($\leq 0.5$ nm) does not strongly attenuate the spin current. This can be understood as the HM/Hf(~0.5)/FeCoB structure having a comparable or smaller SML than that of the seemingly simpler HM/FeCoB bilayer. Since our SO PtMn/FeCoB structures appear to have a quite significant SML, we fabricated a Ta(1.5)/PtMn(8)/Hf(0.25)/FeCoB(0.8)/ MgO(1.6)/Ta(1.5) sample (F) to determine if an ultra-thin Hf insertion layer could be efficacious in this system for enhancing spin transmission and thus $\xi_{DL}$. This sample also exhibits PMA without any high-temperature annealing, and using the HR method we measured an exceptionally high damping-like spin torque efficiency $\xi_{DL} = 0.24 \pm 0.03$. Considering that because of the spin back flow effect, not all of the spin current generated within the PtMn will act on the FM[24,26], this result indicates that the internal spin Hall ratio is $\theta_{SH}^{PtMn} > 0.24$. We summarize the anti-damping torques for series (A)-(F) in Table I.

To determine the spin diffusion length $\lambda_s^{PtMn}$ of our PtMn films we then fabricated a set of samples, series (G), with the multilayer stack being Ta(1)/PtMn($t_{PtMn}$)/Hf(0.8)/FeCoB(0.7)/MgO, where $t_{PtMn}$ ranged from 2 to 8 nm. The thicker Hf spacer promotes strong PMA, with an anisotropy field $H_{an} \approx 1 \text{Tesla}$ over the full range of PtMn thicknesses studied without annealing. In Fig. 2a we show the results of the HR measurements of *anti-damping* like effective field per unit applied electric field $\Delta H_{DL} / E$ as a function of $t_{PtMn}$. It can be shown that (see Supplementary Material):

$$\frac{\Delta H_{DL}}{E} = \frac{\sigma_{SH}}{4\pi M_s t_{FM}^{eff}} \frac{G_A}{G_{PtMn} \tanh(t_{PtMn} / \lambda_s^{PtMn}) + G_B} (1 - \text{sech}(t_{PtMn} / \lambda_s^{PtMn})) \quad (1)$$

where $4\pi M_s$ is the magnetization, $t_{FM}^{eff}$ is the effective thickness of the FM layer excluding the dead layer and $d_{PtMn}$ is the thickness of the PtMn, $\sigma_{SH}$ is the spin Hall conductivity of PtMn ($\sigma_{SH} = \sigma_{PtMn} \theta_{SH}^{PtMn} \hbar / (2e)$), $G_{PtMn} \equiv \sigma_{PtMn} / \lambda_s^{PtMn}$ is the spin conductance of PtMn and $G_A$ and $G_B$ are parameters depending on the Hf spacer and spin mixing conductance at the Hf/FeCoB interface.

Figure 2a shows a fit of equation (1) to the series (G) results, which gives a spin diffusion length of PtMn $\lambda_{\mathrm{PtMn}} = 2.1\mathrm{nm}$. Our result is much larger than the value 0.5 nm previously reported[8] from inverse spin Hall effect (ISHE) measurements on NiFe/PtMn . We note that a significant SML layer in the bilayer system due, for example, to reaction of a component of the FM with Mn at PtMn/FM interface can affect the estimation of $\lambda_s^{\mathrm{PtMn}}$ [18]. We also note that Eqn. (1) assumes a constant spin diffusion length that is independent of $t_{\mathrm{PtMn}}$. This is not necessarily the case if the PtMn resistivity $\rho_{\mathrm{PtMn}}$ varies with film thickness over the range that we are employing and the Elliot-Yafet spin scattering mechanism dominates, where $\lambda_s^{\mathrm{PtMn}} \propto 1/\rho_{\mathrm{PtMn}}$. Fig. 2b shows the measured resistivity of the PtMn thin layers as a function of $t_{\mathrm{PtMn}}$, which is clearly not a constant. Considered this effect, we can use a "rescaling" method introduced in Ref.[27] to fit our data in Fig. 2a, which yields $\lambda_s^{\mathrm{PtMn}} = 2.3\mathrm{nm}$ for the bulk spin diffusion length (see Supplementary Information). This analysis yields a spin conductance for PtMn $G_{\mathrm{PtMn}} = 1/(\lambda_s^{\mathrm{PtMn}} \rho_{\mathrm{PtMn}}) = 0.37 \times 10^{15} \Omega^{-1} \mathrm{m}^{-2}$, considerably lower than that reported[27] for Pt, $G_{\mathrm{Pt}} = 1.3 \times 10^{15} \Omega^{-1} \mathrm{m}^{-2}$ (see also references cited in Ref.[27]). This low PtMn spin conductance could be advantageous in reducing the spin back-flow at an ideal (no SML) PtMn/FM interface (see Ref. [24] and references cited therein).

To demonstrate that PtMn can be used as the source of spin-transfer torque for high-efficiency magnetic switching, we performed current-induced switching using a PtMn(4)/Hf(0.8)/FeCoB(0.7)/MgO sample (H) that as-deposited had strong PMA, and that exhibited sharp and abrupt magnetic switching under an out of plane field as shown in Fig. 3a. An in-plane field $H_x \geq 50 \text{Oe}$ and collinear to the current flow was required to deterministically switch the magnetization, which indicates the existence a weak Dzyaloshinskii-Moriya interaction (DMI) at the Hf/FeCoB interface and a reversal process that proceeds by domain nucleation followed by spin-torque-driven domain expansion[28]. A typical current switching loop is shown in Fig. 3b, as obtained with $H_y = 200 \text{ Oe}$. Fig. 3c shows the spin-torque current switching phase diagram of the same sample. Of course the Hf insertion layer removes the possibility of exchange coupling between the PtMn and the FeCoB, which could add an additional and possibly useful aspect to the simple spin torque switching behavior reported here. We will discuss the switching behavior of PtMn/FM structures with PMA elsewhere.

The value $\xi_{DL} = 0.096 \pm 0.03$ that we obtained from our in-plane magnetized SO PtMn/FeCoB samples is quite similar to that previously reported from inverse spin Hall effect and ST-FMR measurements on in-plane magnetized PtMn/Ni$_{80}$Fe$_{20}$ bilayers ($\theta_{SH}^{PtMn} \approx 0.086$).[8,10] Also the value for $\xi_{FL}$ that we obtain for this set of PtMn/FeCoB samples is comparable to that reported in Ref.[10] from the shift of the resonance field due to a DC current applied during the FMR measurement of the PtMn/Ni$_{80}$Fe$_{20}$ system. In strong contrast to that result, both in our RO PtMn/FeCoB samples and in both versions (SO and RO) of the PtMn/Co sample series $\xi_{DL}$ is much higher. This strongly suggests that a significant SML forms when either FeCoB or Ni$_{80}$Fe$_{20}$ is sputter deposited onto PtMn, but that a weaker SML is the result when Co is deposited onto PtMn. For both Co and FeCoB we find that the weakest SML effect occurs when the deposition order is reversed, *i.e.* in the RO samples. We take this as indicating different degrees of undesirable intermixing in the two deposition orders. For the PMA

PtMn/FeCoB/MgO samples that were deposited in the standard order, $\xi_{FL}$ is quite similar to that measured for the SO samples in the case where the FeCoB layers are thicker and hence magnetized in-plane. However by introducing an ultrathin Hf layer between the PtMn and FeCoB layers, which also enhances the PMA, the SML is greatly suppressed and we obtain $\xi_{DL} = 0.24 \pm 0.03$. This sets only a lower bound on the internal spin Hall ratio of the PtMn $\theta_{SH}^{PtMn}$. Since it is reasonable to expect some remnant SML and/or spin backflow effect at this hybrid interface, it is straightforward to speculate that further efforts to engineer the PtMn/FM interface could result in even higher values of the PtMn spin torque efficiency.

The spin diffusion length of PtMn determined in our measurement is also longer than the previously reported value (<1nm)[8]. We tentatively attribute this to the previous study being sensitive to the formation of a SML layer as the PtMn thickness is increased in those PtMn/Ni$_{80}$Fe$_{20}$ bilayers, as has recently been discussed for the Pt/Co case[18,29]. Of course it has to be considered that the PtMn thickness dependence of $\xi_{DL}$ that we observe is due to some thickness dependent change in the electronic properties of the PtMn film rather than a thicker spin diffusion length than previously determined. It is well known that a fairly thick PtMn layer is required to produce the stable antiferromagnetic domains required for exchange biasing of an adjacent FM film. It is not clear however how this AFM stability would act to enhance the spin current that is generated by the electrical current passing the Pt ions, though we notice that there are seemingly contradicted results on the contribution of macroscopic exchange-bias on SHE in IrMn systems[6,11]. In regard to possible structural changes as a function of PtMn thickness our X-ray diffraction studies (Supplementary Information) do not show any obvious crystalline structure changes for the different thicknesses of PtMn used in this study.

We can use the result for the PtMn spin conductance determined here to further examine the nature of the PtMn/FM interfaces we have studied. In the spin pumping theory[30,31] of a well-ordered HM/FM interface there is an enhancement of the magnetic damping that varies as $\Delta\alpha = (\gamma\hbar^2 / 8\pi^2 e^2 M_s t_{FM}^{eff})G_{eff}^{\uparrow\downarrow}$ where the effective spin mixing conductance $G_{eff}^{\uparrow\downarrow} \equiv G^{\uparrow\downarrow} / (1 + 2G^{\uparrow\downarrow} / G_{HM})$, and $G^{\uparrow\downarrow}$ is the spin mixing conductance of the interface, assuming $\mathrm{Re}\, G^{\uparrow\downarrow} \gg \mathrm{Im}\, G^{\uparrow\downarrow}$ (see Supplementary Information). In all four IPM PtMn/FM series studied, the measurement of $\Delta\alpha(t_{FM}^{-1})$ yielded $G_{eff}^{\uparrow\downarrow} > 0.7\times10^{15}\,\Omega^{-1}\mathrm{m}^{-2}$ (Supplementary Information). With $G_{PtMn} = 0.37\times10^{15}\,\Omega^{-1}\mathrm{m}^{-2}$, this results in an unphysical (negative) value for $G^{\uparrow\downarrow}$, which means that there must be a significant SML at the PtMn/FM interface and/or a non-ideal damping enhancement at the other FM interface[24], neither of which are taken into account in the standard spin pumping theory. (We note that even if we use the previously reported results for PtMn[8] $\rho_{PtMn} = 164\,\mu\Omega\bullet\mathrm{cm}$ and $\lambda_s^{PtMn} = 0.5\mathrm{nm}$ to determine the PtMn spin conductance the $\Delta\alpha(t_{FM}^{-1})$ measurements still yield a negative result for $G^{\uparrow\downarrow}$.)

In summary, depending on the protocol for forming the PtMn/FM interface we have obtained very high *anti-damping* spin torque efficiencies $\xi_{DL}$ from the spin Hall effect in PtMn, with the highest value $\xi_{DL} = 0.24 \pm 0.03 = T_{int} \cdot \theta_{SH}^{PtMn}$ being obtained with a PtMn/Hf(0.25)/FeCoB multilayer, where $T_{int}$ is the net interface spin transparency of that particular system. Assuming that the intrinsic spin Hall effect dominates in PtMn this result provides a lower bound for the spin Hall conductivity of PtMn $\sigma_{SH}^{PtMn} = (\xi_{DL} / T_{int}) \cdot \sigma_{PtMn} > 1.5\times10^5 (2e / \hbar)\Omega^{-1}\mathrm{m}^{-1}$, since $T_{int} < 1$. This can be compared to the lower bound that has been established for Pt, $\sigma_{SH}^{Pt} > 2.8\times10^5 (2e / \hbar)\Omega^{-1}\mathrm{m}^{-1}$ from recent measurements of $\xi_{DL}$ in the PMA Pt/Co system[27].

Refinements that yield a higher $T_{\text{int}}$ for PtMn/FM interfaces will result in even higher $\xi_{DL}$. We conclude that PtMn in particular and likely other binary Pt compounds in general are very promising candidates as spin current sources and detectors in spintronics applications in both IPM and PMA systems provided that the interface can be engineered to have a high spin transparency.

## Methods

### Sample fabrication

All samples in this work were prepared by direct current (DC) sputtering (with RF magnetron sputtering for the MgO layer) in a deposition chamber with a base pressure $< 8 \times 10^{-8}$ Torr. The DC sputtering condition is 2mTorr Ar pressure, 30 watts power and low deposition rates (Ta: $0.0142$ nm/s, PtMn: $0.0189$ nm/s, FeCoB: $0.0064$ nm/s, Co: $0.0066$ nm/s). The PtMn alloy is deposited from a 2-inch planar $Pt_{50}Mn_{50}$ target. We utilized a Ta seeding layer as a template for smoothing the growth of the PtMn for all the standard stacking order samples. All samples have a Ta(1.5) top layer to provide an oxidized protection layer for the stack. We annealed the samples twice at $115\ ^\circ$C for 1 min as part of the photolithography process.

**Acknowledgements**

This research was supported by the Office of Naval Research, and by the NSF/MRSEC program (DMR-1120296) through the Cornell Center for Materials Research. We also acknowledge support from the NSF (Grant No. ECCS-1542081) through use of the Cornell Nanofabrication Facility/National Nanofabrication Infrastructure Network.

# Figure 1

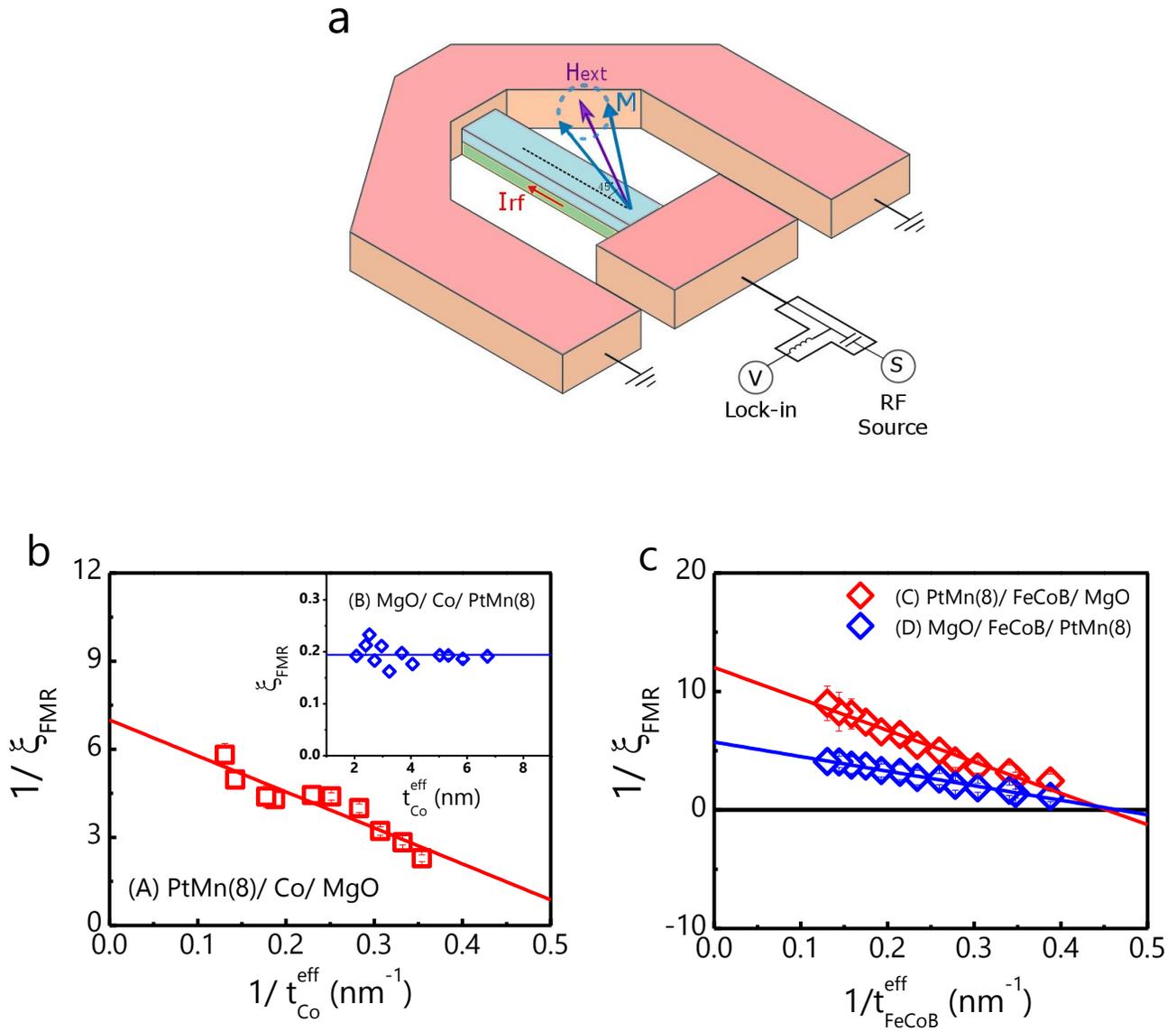

Figure 1: ST-FMR measurement on PtMn/Co and PtMn/FeCoB bilayer samples. **a.** Schematic of ST-FMR measurement. **b,** The inverse of the ST-FMR measured spin torque efficiency, $1/\xi_{FMR}$, as a function of the inverse of the effective thickness for the Co series (A) samples ( red squares). Inset: $\xi_{FMR}$ as a function of $t_{Co}^{eff}$ for series (B) samples (blue squares). **c,** $1/\xi_{FMR}$ as a function of $1/t_{FeCoB}^{eff}$ for the series (C) ( red squares) and series (D) (blue squares) samples.

# Figure 2

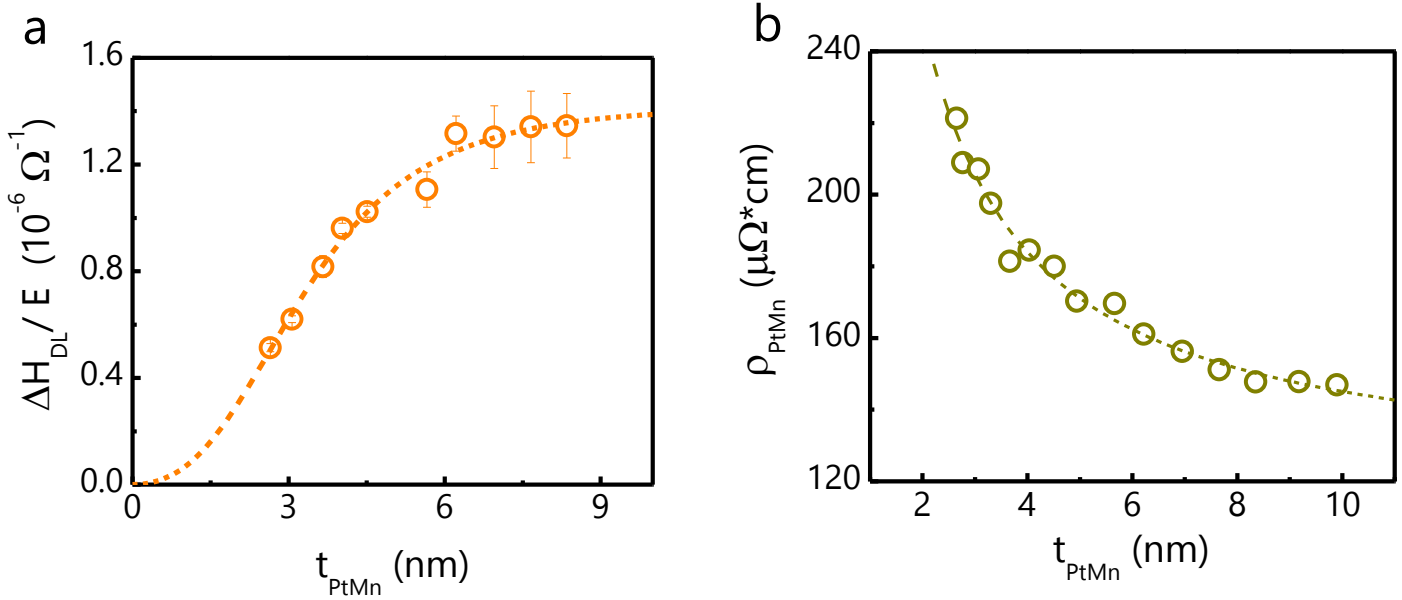

Figure 2: Spin diffusion length and resistivity measurement. **a**, Damping-like effective field per unit applied electric field for the series (G) samples as a function of PtMn thickness $t_{PtMn}$. **b**, Average resistivity of different thickness of PtMn as a function of $t_{PtMn}$. The dash line is a fit to the empirical function $\rho_0 + \rho_s / t_{PtMn}$ to the data, where $\rho_0$ and $\rho_s$ are represent the bulk and interfacial scattering contributions to the resistivity.

# Figure 3

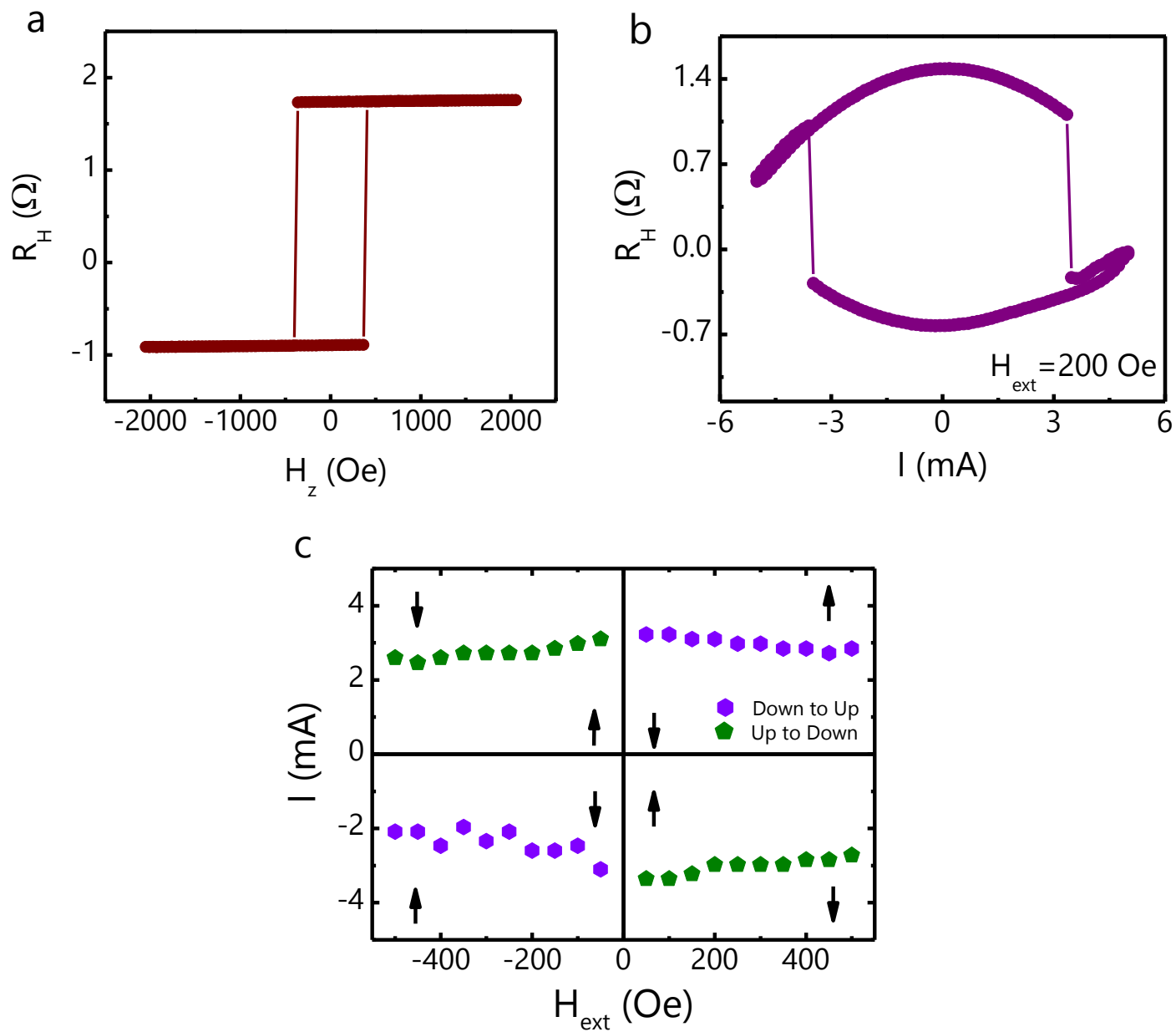

Figure 3: Field and current-induced switching. **a**, Magnetic field switching of a series (H) sample with the field perpendicular to the sample plane. **b**, Current-induced switching of same sample with an external magnetic field (200 Oe) applied in-plane along the current direction. **c**, Phase diagram of the current-induced switching.

Table I

| Samples | Layer Structure | Stack order | $\xi_{DL}$ | Anisotropy |
|---------|-----------------|-------------|------------|------------|
| (A) | $\parallel$Ta(1.5)/PtMn(8)/Co($t_{Co}$)/MgO(1.6)/Ta(1.5) | SO | $0.16 \pm 0.01$ | IPM |
| (B) | $\parallel$MgO(1.6)/Co($t_{Co}$)/PtMn(8)/MgO(1.6)/Ta(1.5) | RO | $0.19 \pm 0.02$ | IPM |
| (C) | $\parallel$Ta(1.5)/PtMn(8)/FeCoB($t_{FeCoB}$)/MgO(1.6)/Ta(1.5) | SO | $0.096 \pm 0.03$ | IPM |
| (D) | $\parallel$MgO(1.6)/FeCoB($t_{FeCoB}$)/PtMn(8)/MgO(1.6)/Ta(1.5) | RO | $0.174 \pm 0.04$ | IPM |
| (E) | $\parallel$Ta(1.5)/PtMn(8)/FeCoB(0.8)/MgO(1.6)/Ta(1.5) | SO | $0.11 \pm 0.02$ | PMA |
| (F) | $\parallel$Ta(1.5)/PtMn(8)/Hf(0.25)/FeCoB(0.8)/MgO(1.6)/Ta(1.5) | SO | $0.24 \pm 0.03$ | PMA |

Table I. A summary of the parameters of the different samples of samples in this study: Here $\parallel$ represents the Si/SiO$_2$ substrate,

SO(RO) means "standard order" ("reverse order") of stack growth, $\xi_{DL}$ is the anti-damping spin torque efficiency of each samples,

and IPM (PMA) means the sample is in-plane (out-of-plane) magnetized.

# Supplementary Material

## Strong Spin Hall Effect in the Antiferromagnet PtMn

Yongxi Ou[1], Shengjie Shi[1], D. C. Ralph[1,2], and R. A. Buhrman[1,*]

[1]Cornell University, Ithaca, New York 14853, USA

[2]Kavli Institute at Cornell, Ithaca, New York 14853, USA

**Contents**

## S1. ST-FMR formula for systems with a non-negligible field-like term

We use the definition utilized in Ref. [1, 2]:

$$\xi_{FMR} = \frac{S}{A}\left(\frac{e}{\hbar}\right) 4\pi M_s t_{FM}^{eff} d_{NM} \sqrt{1+\left(4\pi M_{eff}/H_0\right)} \qquad (1)$$

where $e$, $\hbar$, $4\pi M_s$, $t_{FM}^{eff}$, $d_{NM}$, $H_0$ represent the electron charge, the Planck constant, the magnetization, the effective thickness of FM layer, the thickness of NM layer, the effective demagnetization field of FM layer, and the ferromagnetic resonance field, respectively. $S$ is the symmetric component of the ST-FMR resonance about the resonant field and $A$ is the antisymmetric component. For a HM/FM bilayer system in which there is a damping-like torque $\tau_{DL}$ acting on the FM, but no field-like torque $\tau_{FL}$, $\xi_{FMR}$ is simply equal to the anti-damping spin torque efficiency $\xi_{DL}$. However if the field-like spin torque efficiency $\xi_{FL}$ is not negligible, one can express $S(A)$ as:

$$S = \frac{\hbar}{2e}\frac{\xi_{DL}J_e^{rf}}{4\pi M_s t_{FM}^{eff}} \qquad (2)$$

$$A = \left(H_T + H_{Oe}\right)\sqrt{1+\left(4\pi M_{eff}/H_0\right)} = \left(\frac{\hbar}{2e}\frac{\xi_{FL}J_e^{rf}}{4\pi M_s t_{FM}^{eff}} + \frac{J_e^{rf}d_{NM}}{2}\right)\sqrt{1+\left(4\pi M_{eff}/H_0\right)} \qquad (3)$$

where $J_e^{rf}$, $H_T \propto \tau_{DL}$, $H_{Oe}$ are the electric current density, field-like effective field and the Oersted field, respectively. Combining Eqs. (1)-(3), we have:

$$\frac{1}{\xi_{FMR}} = \frac{1}{\xi_{DL}}\left(1+\frac{\hbar}{e}\frac{\xi_{FL}}{4\pi M_s t_{FM}^{eff} d_{NM}}\right) \qquad (4)$$

As long as $\xi_{FL}$ is independent of FM thickness, then $1/\xi_{FMR}$ will be a linear function of $1/t_{FM}^{\text{eff}}$, from which the intercept and slope allow a determination of $\xi_{DL}$ and $\xi_{FL}$. We use Eq. (4) to fit the ST-FMR data for IPM series (A)-(D) samples in the main text.

## S2. Determination of PMA in Ta/PtMn(8)/FeCoB( t$_{FeCoB}$ )/MgO samples

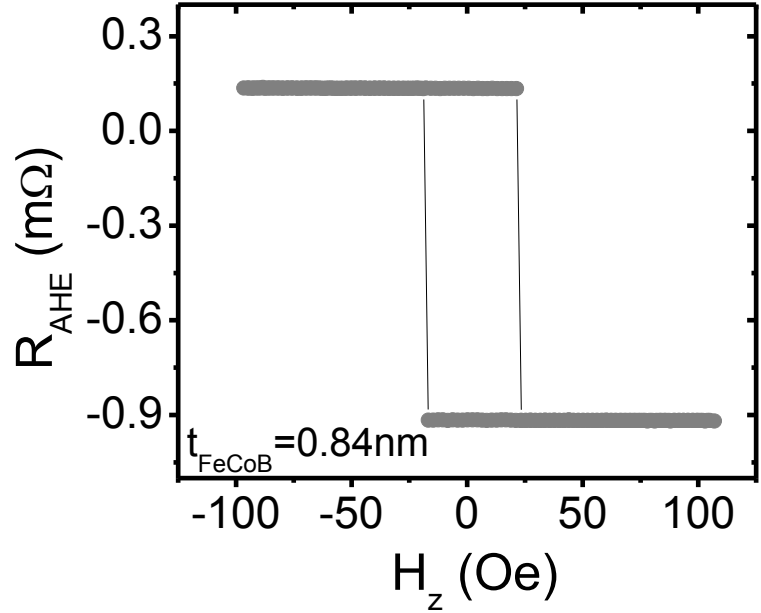

Figure S2: Anomalous hall measurement. Anomalous Hall resistance of sample

Ta(1.5)/PtMn(8)/FeCoB($t_{FeCoB}$ )/MgO(1.6)/Ta(1.5) when $t_{FeCoB}$ is equal to 0.84 nm.

We determined the thickness range for FeCoB that yields PMA for the multilayer

Ta(1.5)/PtMn(8)/FeCoB($t_{FeCoB}$ )/MgO(1.6)/Ta(1.5) by performing anomalous Hall resistance

measurements under an external magnetic field applied perpendicular to the film plane. A typical

sample is shown in Fig. S2. We found the PMA thickness range for this sample to be

$0.6$ nm $< t_{FeCoB} < 1.0$ nm, with the strongest PMA occuring when $t_{FeCoB} \sim 0.77$ nm.

**S3. Spin diffusion length measurement of PtMn**

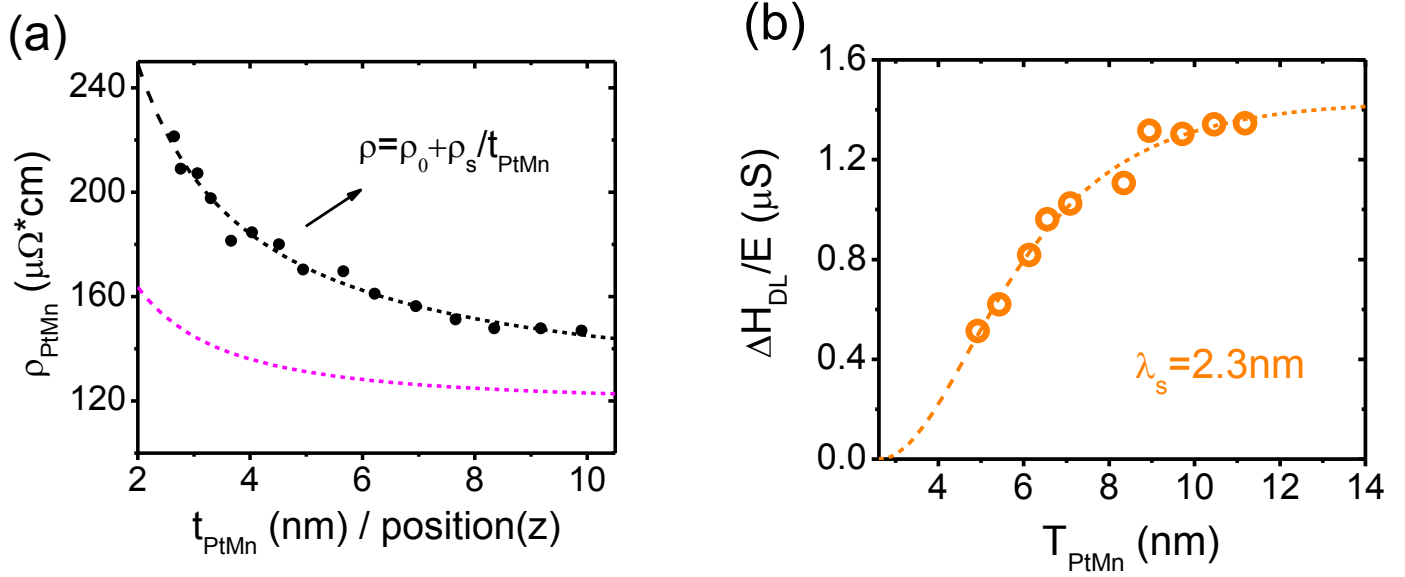

Figure S3: Resistivity and spin diffusion length measurement. **a,** The measured average resistivity and local resistivity of PtMn in series (B) samples. The black dashed line is a fit to the average resistivity while purple dashed line represents the calculated local resistivity. **b,** Damping-like effective spin-torque field per unit applied electric field as a function of the effective thickness of PtMn. The dash line represents a fit using Eq. (9).

In a diffusive model for spin transport within a PtMn($t_{PtMn}$)/Hf/FeCoB trilayer, the damping-like spin torque efficiency should depend on the PtMn thickness $t_{PtMn}$ as[24]:

$$\xi_{DL} = \theta_{SH}(1 - \text{sech}(\frac{t_{PtMn}}{\lambda_{PtMn}})) \frac{G^*}{G^* + G_{PtMn}\tanh(t_{PtMn}/\lambda_{PtMn})} \text{sech}(\frac{t_{Hf}}{\lambda_{Hf}}) \frac{2\,\text{Re}(G^{\uparrow\downarrow})}{2\,\text{Re}(G^{\uparrow\downarrow}) + G_{Hf}} \qquad (5)$$

Here $t_{Hf}$, $\lambda_{Hf}$, $G_{Hf}$ are the thickness, the spin diffusion length and spin conductance of Hf (with $G_{Hf} = 1/(\rho_{Hf}\lambda_{Hf})$), and

$$G^* = G_{Hf}\coth(\frac{t_{Hf}}{\lambda_{Hf}}) \frac{G_{Hf}\tanh(\frac{t_{Hf}}{\lambda_{Hf}}) + 2\,\text{Re}(G^{\uparrow\downarrow})}{G_{Hf}\coth(\frac{t_{Hf}}{\lambda_{Hf}}) + 2\,\text{Re}(G^{\uparrow\downarrow})} \qquad (6)$$

We define the anti-damping spin torque effective field $\Delta H_{DL}$ such that[32]:

$$\xi_{DL} = \left(\frac{2e}{\hbar}\right) 4\pi M_s t_{FM}^{\text{eff}} \left(\frac{\Delta H_{DL}}{J_e}\right) \qquad (7)$$

Combining Eqs. (5) and (7) gives:

$$\frac{\Delta H_{DL}}{E} = \frac{\sigma_{SH}}{4\pi M_s t_{FM}^{\text{eff}}} \frac{G_A}{G_{PtMn}\tanh(t_{PtMn}/\lambda_{PtMn}) + G_B}(1 - \text{sech}(t_{PtMn}/\lambda_{PtMn})) \qquad (8)$$

where $G_A$ ($G_B$) is a parameter depending on the Hf spacer and spin mixing conductance at Hf/FeCoB. This is Eq. (1) in the main text.

We used Eq. (8) to fit the data in Fig. 2a (main text) to obtain an effectives spin diffusion length of PtMn = 2.1 nm. However, Eq. (8) assumes that the spin diffusion length and the spin Hall ratio in the PtMn layer are both constant, independent of $t_{PtMn}$. This is not necessarily the case if the PtMn resistivity $\rho_{\text{PtMn}}$ varies with film thickness over the range that we are employing. In this case if the spin relaxation in PtMn is dominated by the Elliot-Yafet scattering mechanism where $\rho_{\text{PtMn}} \times \lambda_s^{\text{PtMn}}$ is the constant quantity, $\lambda_s^{\text{PtMn}}$ will vary with $t_{\text{PtMn}}$. In addition, if the spin Hall effect in PtMn is dominant by the intrinsic process, rather than by extrinsic skew-scattering, the spin Hall ratio is not constant but varies linearly with resistivity, $\theta_{SH}^{\text{PtMn}} = (\hbar / 2e)\rho_{\text{PtMn}} \times \sigma_{SH}^{\text{PtMn}}$, where $\sigma_{SH}^{\text{PtMn}}$ is the spin Hall conductivity and is expected to be constant when the resistivity is altered by a change in the elastic and quasi-elastic scattering rates within the material. The resistivity of our PtMn thin films as averaged over the thickness of each film is shown in Fig. 2b (main text) as a function of $t_{\text{PtMn}}$, and there is clearly a substantial variation from 2 nm to 10 nm. In a recent report on the thickness dependent spin Hall properties of Pt thin films, our group has proposed a "rescaling" method for dealing with this effect in the case of Elliot-Yafet spin scattering and the intrinsic spin Hall effect regime[27]. Here we use a similar treatment to our PtMn data. First we measured the resistivity of PtMn layer in sample (F) Ta(1)/PtMn($t_{\text{PtMn}}$)/Hf(0.8)/FeCoB(0.7)/MgO by comparing its resistance to a control sample without PtMn layer: Ta(1)/ Hf(0.8)/FeCoB(0.7)/MgO. The resistance and thus resistivity of PtMn layer can be determined from the parallel resistance model. The result is shown in Fig. 3(a) in the main text and in Fig. S4(a) with black dots. We used the form $\rho(t_{\text{PtMn}}) = \rho_0 + \rho_s / t_{\text{PtMn}}$ to fit the data, in which $\rho_0$ and $\rho_s$ represent the bulk and interfacial scattering contribution to the average resistivity $\rho(t_{\text{PtMn}})$ for a certain thickness of PtMn. The fit (black dashed line) in Fig. S3(a) gives $\rho_0 = 119 \ \mu\Omega\text{cm}$ and $\rho_s = 2.6 \times 10^{-5} \ \mu\Omega\text{cm}^2$. Then by considering that a certain thickness of PtMn layer consists of slices of PtMn thin films with varying resistivity, we calculated the "local resistivity" $\rho(z)$ of PtMn as $\rho(z) = (\rho_0 + \rho_s / z)^2 / (\rho_0 + 2\rho_s / z)$, shown as the purple dashed line in Fig. S3(a). Based on $\rho(z)$, we can transform the thickness of PtMn $t_{\text{PtMn}}$ into an effective thickness $T_{\text{PtMn}}$ with the same spin diffusion length $\lambda_0$ corresponding to the bulk value of resistivity $\rho_0$, and Eq. (8) becomes:

$$\frac{\Delta H_{DL}}{E} = \frac{\sigma_{SH}}{4\pi M_s t_{FM}^{\text{eff}}} \frac{G_A}{G_{\text{PtMn}} \tanh((T_{\text{PtMn}} - T_0)/\lambda_0) + G_B}(1 - \text{sech}((T_{\text{PtMn}} - T_0)/\lambda_0)) \qquad (9)$$

The fit using Eq. (9) gives $\lambda_0 = 2.3\ \text{nm}$, shown in Fig. S3(b).

**S4. X-ray diffraction measurement**

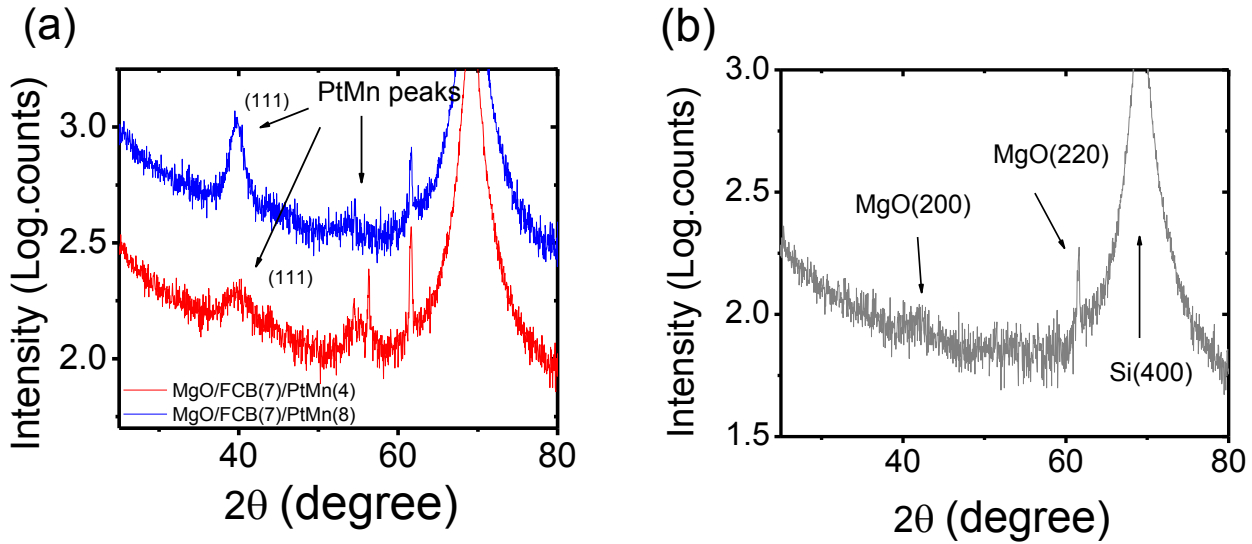

Figure S4: X-ray diffraction measurement. XRD pattern for **a,** sample

MgO(1.6)/FeCoB(7)/PtMn(4)/MgO(1.6)/Ta(1.5) (red) and

MgO(1.6)/FeCoB(7)/PtMn(8)/MgO(1.6)/Ta(1.5) (blue), and **b,** a reference sample

MgO(1.6)/FeCoB(7)/MgO(1.6)/Ta(1.5). The signals in (a) have been shifted in the y-direction to allow for comparison.

We obtained x-ray diffraction spectra from samples

MgO(1.6)/FeCoB(7)/PtMn(4,8)/MgO(1.6)/Ta(1.5), as shown in Fig. S4 (a). The main PtMn

(111) peak is observed at around $2\theta \sim 40°$. With the thinner PtMn layer (4 nm), the peak is

lower and broader in comparison to that of the thicker PtMn (8 nm) layer. This is consistent with the thicker PtMn sample having somewhat larger crystalline domains. There is also a secondary peak centered at $2\theta \sim 55°$, which we have not identified but does appear to arise from the PtMn.

Fig. S4(b) shows the x-ray diffraction measured for a reference sample

MgO(1.6)/FeCoB(7)/MgO(1.6)/Ta(1.5), which has the same stacking order and structure as sample (D) without the PtMn layer.

## S5. Damping measurement and spin mixing conductance calculation

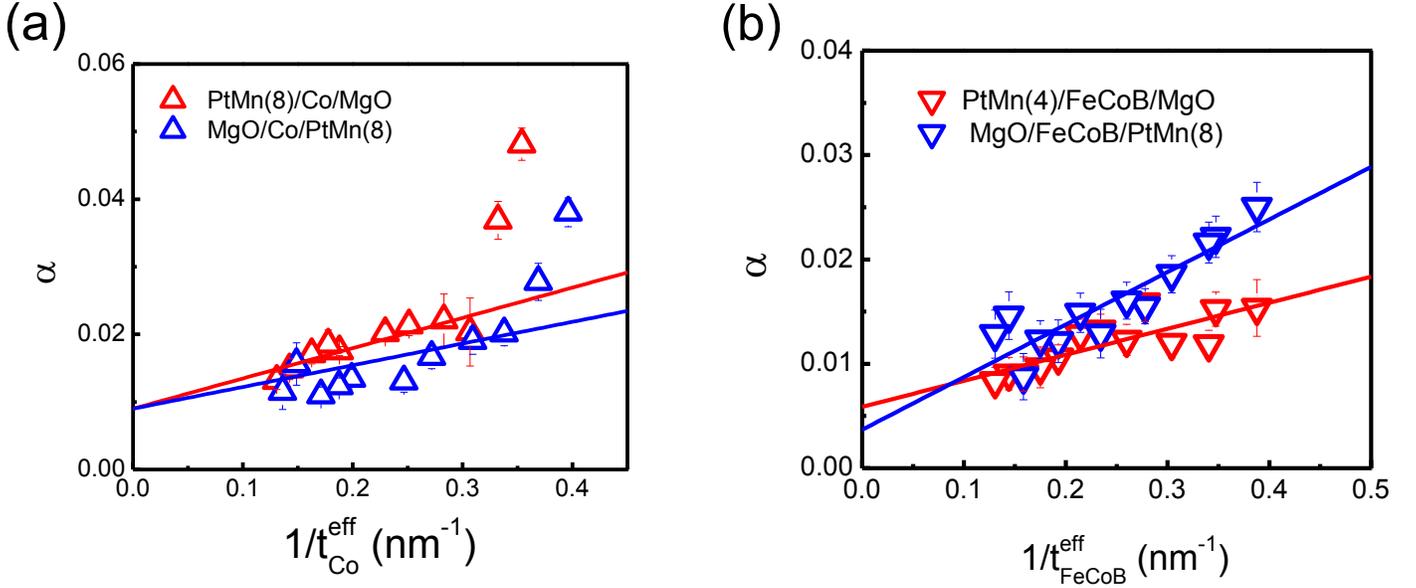

Figure S5: Damping measurements: a,b, Enhanced damping as a function of the inverse of the effective thickness for Co samples SO series (A) (red triangles) and RO series (B) (blue triangles), and FeCoB samples SO series (C) (red triangles) and RO series (D) (blue triangles).

According to the conventional spin pumping theory[30,31], by measuring the enhancement of the damping constant $\Delta\alpha$ as a function of the FM layer thickness, as shown in Fig. 5, and provided that there are no other interfacial contributions to the damping, such as SML or a magnetic "dead layer," one can determine the interfacial spin-mixing conductance:

$$\Delta\alpha = \alpha - \alpha_0 = \frac{\gamma\hbar}{4\pi M_s t_{FM}^{\text{eff}}} g_{\text{eff}}^{\uparrow\downarrow} = \frac{\gamma\hbar^2}{8\pi^2 e^2 M_s t_{FM}^{\text{eff}}} G_{\text{eff}}^{\uparrow\downarrow} \qquad (10)$$

Here $\alpha_0$ and $\gamma = 1.76 \times 10^{11} \text{ s}^{-1}\text{T}^{-1}$ represent the intrinsic damping constant and the gyromagnetic ratio of the FM layer. $g_{\text{eff}}^{\uparrow\downarrow} (G_{\text{eff}}^{\uparrow\downarrow})$ is the so-called interfacial spin-mixing conductance, which is related to the spin current transparency at the PtMn/FM interface[24,26] (with $G_{\text{eff}}^{\uparrow\downarrow} = (e^2 / h) g_{\text{eff}}^{\uparrow\downarrow}$).

However, if additional spin attenuation occurs in the interfaces of the FM layer, that would contribute to $g_{\text{eff}}^{\uparrow\downarrow}$, making it larger than the value given only by spin pumping[24], leading to an over-estimate of the true interfacial spin-mixing conductance. Fitting the data in Fig. S5 gives the apparent $g_{\text{eff}}^{\uparrow\downarrow} (G_{\text{eff}}^{\uparrow\downarrow})$ as summarized in the following table for samples (A)-(D):

| Sample | Apparent $g_{\text{eff}}^{\uparrow\downarrow}$ (nm$^{-2}$) | Apparent $G_{\text{eff}}^{\uparrow\downarrow} = (e^2 / h) g_{\text{eff}}^{\uparrow\downarrow}$ (10$^{15}\,\Omega^{-1}\text{m}^{-2}$) |
|---|---|---|
| A | 40.8 | 1.58 |
| B | 33.4 | 1.29 |
| C | 18.8 | 0.72 |
| D | 39.4 | 1.53 |

The bare spin mixing conductance $G^{\uparrow\downarrow}$ of the interface can be expressed as[24]:

$$G^{\uparrow\downarrow} = \frac{G_{\text{eff}}^{\uparrow\downarrow}}{1 - 2\dfrac{G_{\text{eff}}^{\uparrow\downarrow}}{G_{\text{PtMn}}}} \qquad (11)$$

If we use the result $G_{\text{PtMn}} = 1/(\rho_{\text{PtMn}} \lambda_{\text{PtMn}}) = 0.37 \times 10^{15}\ \Omega^{-1}\text{m}^{-2}$ (from section S3 above), one can see that $G^{\uparrow\downarrow}$ will have to be negative for all four samples, which is unphysical. This is also the case if we use the measured values $\rho_{\text{PtMn}} = 164\ \mu\Omega\text{cm}$ and $\lambda_{\text{PtMn}} = 0.5\ \text{nm}$ from previous PtMn research[8]. This suggests again that the measured $g_{\text{eff}}^{\uparrow\downarrow}$ cannot be explained by the conventional bilayer spin pumping model and extra interfacial spin attenuation factors (not related to injection of spin current into the HM) have to be included to account for the large measured values of $g_{\text{eff}}^{\uparrow\downarrow}$.